\documentclass[prl,twocolumn,superscriptaddress,showpacs,preprintnumbers]{revtex4}

\usepackage{graphicx}
\usepackage{bm}
\usepackage[usenames,dvipsnames]{color}

\usepackage{amsmath}

\begin{document}

\title{Near-unity third-harmonic circular dichroism driven by quasi-BIC in asymmetric silicon metasurfaces}

\author{Marco Gandolfi}
\email{marco.gandolfi@ino.cnr.it}
\affiliation{CNR-INO (National Institute of Optics), Via Branze 45, Brescia, Italy}
\affiliation{Department of Information Engineering, University of Brescia, Via Branze 38, Brescia, Italy}

\author{Andrea Tognazzi}
\affiliation{CNR-INO (National Institute of Optics), Via Branze 45, Brescia, Italy}
\affiliation{Department of Information Engineering, University of Brescia, Via Branze 38, Brescia, Italy}

\author{Davide Rocco}
\affiliation{CNR-INO (National Institute of Optics), Via Branze 45, Brescia, Italy}
\affiliation{Department of Information Engineering, University of Brescia, Via Branze 38, Brescia, Italy}

\author{Costantino De Angelis}
\affiliation{CNR-INO (National Institute of Optics), Via Branze 45, Brescia, Italy}
\affiliation{Department of Information Engineering, University of Brescia, Via Branze 38, Brescia, Italy}

\author{Luca Carletti}
\email{luca.carletti@unibs.it}
\affiliation{CNR-INO (National Institute of Optics), Via Branze 45, Brescia, Italy}
\affiliation{Department of Information Engineering, University of Brescia, Via Branze 38, Brescia, Italy}

\begin{abstract}
We use numerical simulations to demonstrate third-harmonic generation with near-unity nonlinear circular dichroism (CD) and high conversion efficiency ($ 10^{-2}\ \text{W}^{-2}$) in asymmetric Si-on-SiO$_2$ metasurfaces.
The working principle relies on the spin-selective excitation of a quasi-bound state in the continuum, characterized by a very high ($>10^5$) quality-factor. 
By tuning multi-mode interference with the variation of the metasurface geometrical parameters, we show the possibility to control both linear CD and nonlinear CD.
Our results pave the way for the development of all-dielectric metasurfaces for nonlinear chiro-optical devices with high conversion efficiency.

\end{abstract}
                              
\maketitle
\begin{figure}[t]
    \centering
    \includegraphics[width=0.25\textwidth]{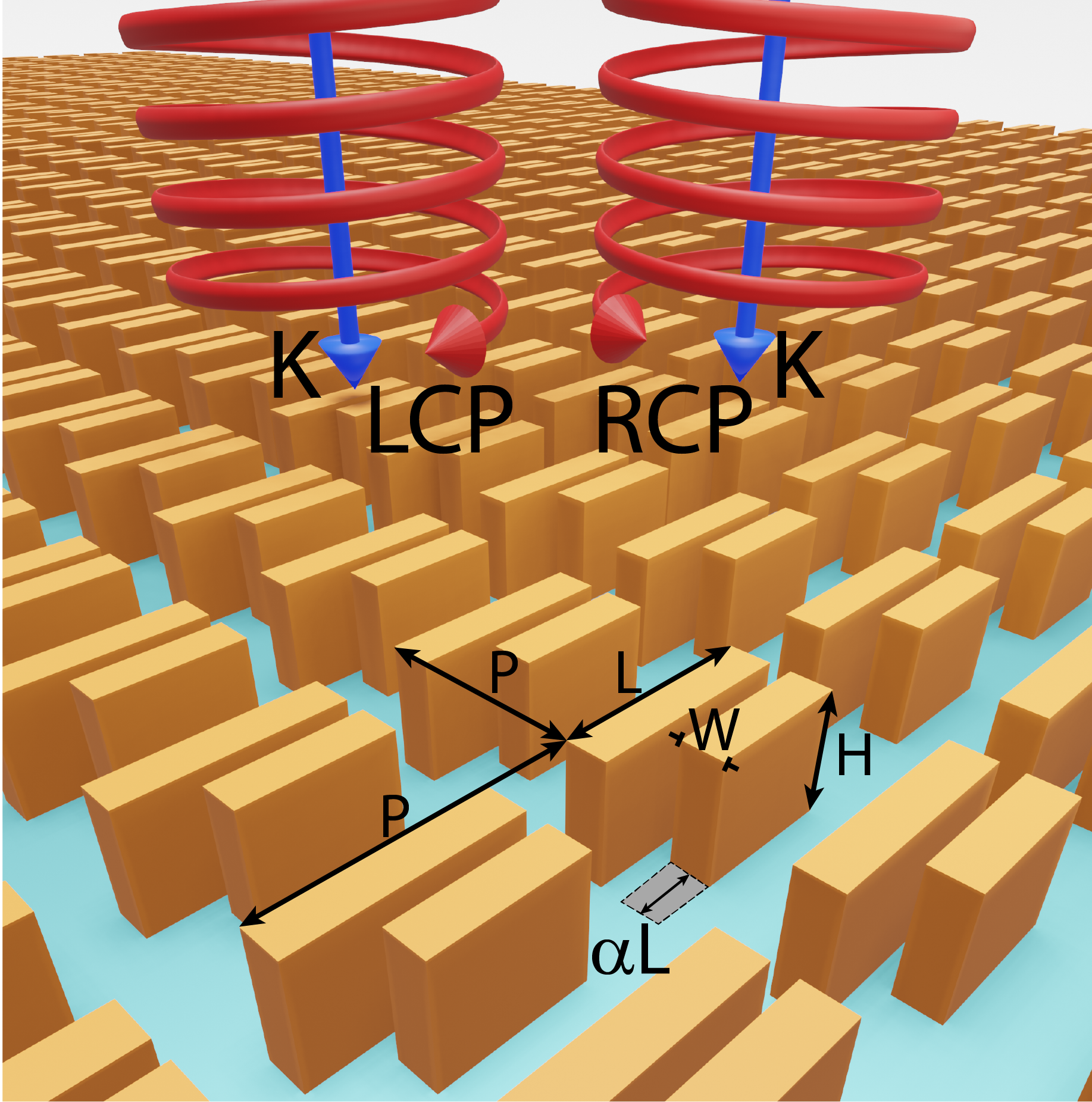}
    \caption{Sketch of the MS with the relevant geometrical parameters and of the illumination scheme with LCP and RCP light.}
    \label{fig:sketch}
\end{figure}

Polarization is one of the most fundamental properties of electromagnetic radiation.
The capability to measure, manipulate, and control the polarization state of a light beam is essential for a huge range of applications such as classical and quantum optical communications \cite{wang2012terabit,wang2015quantum,virte2013deterministic}, biological and chemical sensing \cite{barron2009molecular,estephan2010sensing,pellegrini2018surface}, imaging and holography \cite{ni2013metasurface,lee2020metasurfaces}, spectropolarimetry and metrology \cite{sterzik2012biosignatures,salvail2013full}.
In this context, optical rotation (OR) - rotation of the polarization axis of linearly polarized waves - and circular dichroism (CD) - difference in the absorption/reflectivity/transmittivity spectra of left-circularly polarized (LCP) and right-circularly polarized (RCP) light - are of paramount importance. Although these chiro-optical phenomena are extremely weak in natural materials, the recent development of chiral metamaterials and metasurfaces (MSs) \cite{wang2016optical,valev2013chirality,mun2020electromagnetic,hentschel2017chiral,keren2016nonlinear} has demonstrated giant OR and CD enhancement suitable for chiral biosensing and imaging \cite{hentschel2017chiral}, polarization conversion \cite{zhao2012twisted}, negative refraction \cite{pendry2004chiral}, spin-controlled wavefront shaping \cite{ye2016spin}, and chiral detectors \cite{li2015circularly}.\\
\indent Most chiral metamaterials demonstrated so far are made of three-dimensional (3D) metallic nanostructures \cite{wang2016optical,valev2013chirality,mun2020electromagnetic,hentschel2017chiral,Fasold2018,tanaka2020chiral}.
However, the complexity of the associated fabrication technology and optical absorption have hindered their application. Although all-dielectric chiral metamaterials have been recenlty demonstrated \cite{tanaka2020chiral}, two-dimensional (2D) or planar metamaterials - i.e. MSs - are highly attractive due to compatibility with planar fabrication technologies. Despite a lower chiro-optical response due to inherent reflection symmetry, planar structures sustaining Fano resonances can exhibit strong non-reciprocal CD due to modal interference \cite{Hopkins2016}.
Thus, the recent insurgence of dielectric nanostructures and MSs enabled the realization of lossless planar structures with strong chiro-optical effects \cite{kuznetsov2012magnetic,anthur2020continuous,gomez2019all,gorkunov2018chiral,ullah2018chiral,zhu2018giant,solomon2018enantiospecific,hu2019high,wu2014spectrally,ma2018all,gorkunov2020metasurfaces}.
Chiral dielectric MSs are cornerstone for developing efficient integrated components enabling the measurement of the polarization state of light which is an essential functionality for many applications \cite{li2015circularly,gorkunov2020metasurfaces,tanaka2020chiral,overvig2021chiral}.
\\
\indent Over the past decade, investigations of nonlinear optical effects in chiral metamaterials and MSs have grown in interest \cite{wang2016optical,li2017nonlinear,ren2012giant,chen2016giant,tang2020nano,kimdielectric,kim2020giant} due to the possibility to enhance the chiro-optical response. Nonlinear CD, i.e. strong difference in second-harmonic generation (SHG) or third-harmonic generation (THG) conversion efficiency between LCP and RCP pump, opens to new possible applications such as nonlinear holography, nonlinear wavefront control, and chiral sensors. Although near-unity nonlinear CD has been demonstrated \cite{wang2016optical,li2017nonlinear,ren2012giant,chen2016giant,tang2020nano}, the conversion efficiency and pump intensity have been limited by the employment of metallic structures. Recently, enhanced nonlinear CD with high conversion efficiency has been demonstrated in an hybrid multiple-quantum-well structure \cite{kim2020giant,makarov2017efficient}.
However, optical absorption still limits the pump power that these structures can sustain. Albeit mainly plasmonic and hybrid structures have been investigated so far, dielectric resonators demonstrated considerably higher SHG and THG conversion efficiencies \cite{pertsch2020nonlinear}. In particular, concepts inspired by the bound state in the continuum (BIC) physics are providing a new framework to enhance nonlinear optical phenomena at the nanoscale \cite{carletti2018giant,koshelev2020subwavelength,carletti2019high,koshelev2019nonlinear,anthur2020continuous,liu2019high,overvig2021chiral}. 
Non-plasmonic symmetry-broken structures sustaining quasi-BIC, which were recently shown to display a very important linear CD in the transmittivity \cite{gorkunov2020metasurfaces,overvig2021chiral}, are thus promising to enable both high nonlinear CD and high conversion efficiency \cite{chen2016giant,li2017nonlinear,kim2020giant,kimdielectric,frizyuk2021nonlinear}.\\
\indent In this work, we use numerical simulations to demonstrate symmetry-broken Si MSs supporting a quasi-BIC, the latter enhancing THG conversion efficiency up to $10^{-2}\,\text{W}^{-2}$ and enabling near-unity nonlinear CD ($>99.9\%$).
The high ($>10^5$) quality-factor ($Q$-factor) of our engineered quasi-BIC and the high value of the third-order nonlinear susceptibility ($\chi^{(3)}$) in silicon in the near-infrared spectrum render this approach promising for an efficient THG.
Our results show that the quasi-BIC is selectively excited only for one circular polarization enabling near unity nonlinear CD.\\
The unique opportunity to engineer a high-Q mode in lossless dielectric metasurfaces provides the unprecedented possibility to enhance simultaneously the nonlinear response and the nonlinear CD which, so far, have been hindered by the use of plasmonic metasurfaces.\\
\indent The designed MS is depicted in Fig. \ref{fig:sketch}.
It is composed by periodic unit cells made by two Si blocks on a silica substrate, which is compatible with the silicon-on-insulator material system. The MS periodicity is $P$, whereas the width and the height of each Si block are $W$ and $H$, respectively.
The distance between the two blocks is $t_{gap}$.
To induce the symmetry breaking, the length of one block in the unit cell is reduced by the quantity $\alpha L$, where $L$ is the length of the unperturbed block and $\alpha$ ($0\leq \alpha \leq 1$) is the \textit{asymmetry parameter} (see Fig. \ref{fig:sketch}).
When $\alpha=0$, this MS can sustain symmetry protected BICs. As the symmetry of the unit cell is broken, the BIC evolves into a quasi-BIC due to the opening of a radiation channel to free-space \cite{koshelev2018asymmetric,koshelev2019nonlinear}.
In this work we fix the following geometrical parameters as: $P=700\,\text{nm}$, $W=214\,\text{nm}$, $H=575\,\text{nm}$, and $t_{gap}=116\,\text{nm}$, chosen in order to tune the BIC wavelength in the telecom range.\\
\indent We develop a full-vectorial numerical model in COMSOL Multiphysics \cite{gandolfi2018ultrafast} for the calculation of the optical eigenmodes in the structure (see Appendix for further details).
\begin{figure*}
    \centering
    \includegraphics[width=0.92\textwidth]{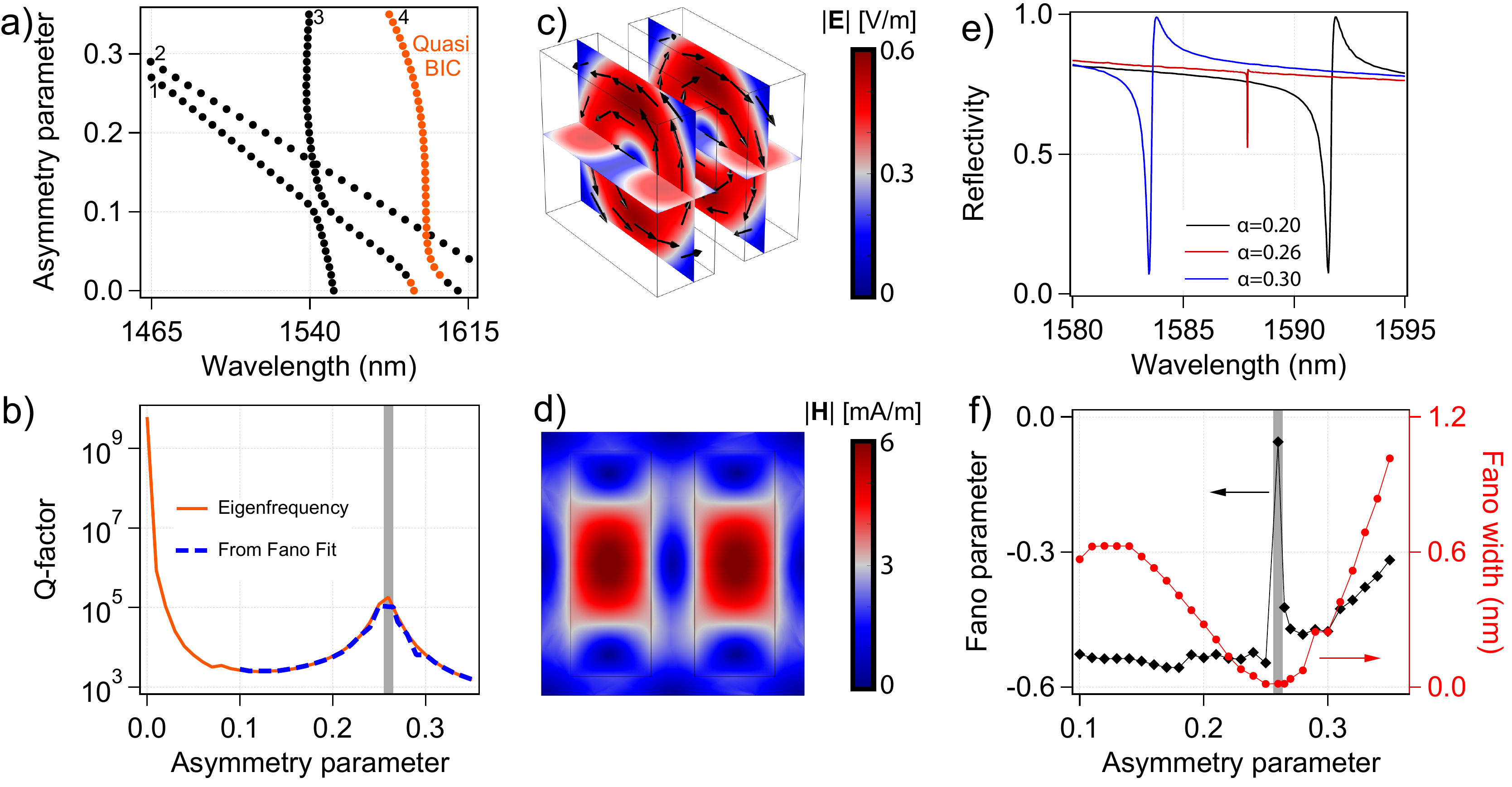}
    \caption{a) Eigenmodes wavelength (horizontal axis) vs asymmetry parameter (vertical axis).
    The numbers in the panel indicate the labels assigned to the bands.
    The quasi-BIC is highlighted with orange markers.
    b) $Q$-factor of the quasi-BIC as a function of the asymmetry parameter. The full orange (dashed blue) line is obtained from the eigenfrequency analysis (from fit of the reflectivity spectra with the Fano formula).
    c) The color map shows the normalized electric field magnitude inside the two Si blocks for the quasi-BIC. The electric field vector is shown with black arrows.
    d) Cross sectional view (over a plane parallel to the substrate-air interface and passing through the blocks centers) of the normalized magnetic field amplitude. The latter two panels refer to the BIC occurring for $\alpha=0$.
    e) Reflectivity spectra for three different asymmetry parameters.
    f) Fano parameter, $q$, (left axis, black) and width, $\gamma$, (right axis, red) vs asymmetry parameter.
    The gray vertical bar in b) and f) highlights the $Q$-factor local maximum.
    All the panels are obtained for $L=600$ nm.}
    \label{fig:modal_analysis}
\end{figure*}
We start our analysis from the case $L=600$ nm and we study how the modes in the MS evolve as the asymmetry parameter increases from $0$ up to $0.35$.
Fig. \ref{fig:modal_analysis}a shows the wavelength, $\lambda$, (horizontal axis) of the MS eigenmodes as a function of the asymmetry parameter (vertical axis).
We highlight in orange the quasi-BIC which is characterized by a much higher $Q$-factor, the latter defined as 
\begin{equation}
Q=\frac{1}{2}\frac{Real(\tilde{\nu})}{Imag(\tilde{\nu})},
\label{eq:Q_wef}
\end{equation}
where $\tilde{\nu}$ is the complex eigenfrequency of the mode.
The $Q$-factor of the quasi-BIC as a function of $\alpha$ is shown in
Fig. \ref{fig:modal_analysis}b (orange curve). For $\alpha=0$, the $Q$-factor diverges towards infinity and observation of the electric and magnetic
field profiles shown in Figs. \ref{fig:modal_analysis}c and \ref{fig:modal_analysis}d reveals that
it is a symmetry protected BIC.
In particular, each block sustains a magnetic dipole mode with opposite momentum that results in an odd-symmetric y-component of the electric field at the MS-air interface.
As $\alpha$ is slightly increased ($0<\alpha\leq0.1$), the BIC evolves into a quasi-BIC due to the opening of a radiation channel to free-space and  $Q\propto\alpha^{-2}$\cite{koshelev2018asymmetric,koshelev2019nonlinear}.\\
\indent For further increase of $\alpha$, we notice that the $Q$-factor shows a local maximum around the critical value $\alpha^*=0.25$.
\indent The origin of the latter can be ascribed to accidental BIC formed due to a simultaneous destructive interference of leakage channels \cite{Hsu2016}, as demonstrated at THz frequencies \cite{Han2021}.\\
\indent The quasi-BIC physics is further investigated by solving the electromagnetic scattering problem with full-vectorial numerical calculations (see Appendix for further details).
The incident light is linearly polarized with the electric filed parallel to the block side with length $L$ and it impinges on the MS from the air side at normal incidence.
The reflectivity as a function of the wavelength is computed for different asymmetry parameters, $\alpha$.
In Fig. \ref{fig:modal_analysis}e, we report the reflectivity for three cases, i.e. $\alpha$ in proximity of the $Q$-factor local maximum (red curve) and on the two flanks (black and blue curves).
The spectra show a marked Fano resonance in proximity of the quasi-BIC wavelength. At the critical value $\alpha^*$ the resonance width shrinks, consistently with the modal $Q$-factor enhancement.\\
\indent We fit the reflectivity (R) spectra with a Fano function \cite{limonov2017fano}:
\begin{equation}
    R(E)=R_0+B\frac{\left(q\frac{\gamma}{2}+E-E_0\right)^2}{\left(\frac{\gamma}{2}\right)^2+\left(E-E_0\right)^2},
    \label{Eq:Fano_function}
\end{equation}
where $E$ is the photon energy, $E_0$ is the resonance energy, $\gamma$ is the resonance full width at half maximum (FWHM), $q$ is the Fano parameter, $R_0$ is an offset, and $B$ is a multiplicative constant.\\
\indent In Fig. \ref{fig:modal_analysis}f we report the Fano parameter, $q$, (left axis, black) and $\gamma$ (right axis, red) as function of the asymmetry parameter, $\alpha$.
Below $\alpha=0.10$ the quasi-BIC is very close to another mode with lower $Q$-factor (see Fig. \ref{fig:modal_analysis}a), thus our fit with a single Fano resonance does not allow a careful reproduction of the reflectivity curve.\\
\indent The dashed blue line in Fig. \ref{fig:modal_analysis}b shows the $Q$-factor retrieved from Fano fit on the reflectivity curves obtained from the full vectorial simulations (calculated as $Q=\lambda/\gamma$).
This is in excellent agreement with the eigenmode analysis results, meaning that the two methods are consistent and reliable.
We observe that the Fano parameter, $q$, is always negative except in proximity of $\alpha^*$, meaning that in the latter situation the quasi-BIC and the continuum phases are shifted by almost $\pi/2$ \cite{limonov2017fano}.\\
\indent The quasi-BIC with high $Q$-factor occurring for $\alpha\sim0.25$ is very interesting for technological applications.
This scenario, corresponding to a MS with a high asymmetry, supporting a Fano resonance, and exhibiting a strong electric field enhancement at the same time, is the perfect playground where nonlinear CD may be observed.
Indeed, the exploitation of a quasi-BIC with high $Q$-factor is beneficial to enhance the electric field within the structure.
Temporal coupled-mode theory indicates that the coupling efficiency of the incident energy to the resonator scales as the inverse of the $Q$-factor while the electric field is enhanced due to the resonant response of the resonator and scales as $Q^2$ \cite{maier2006plasmonic}. As a result, the electric field enhancement inside the resonator is proportional to $Q$, thus the quasi-BIC is expected to yield a strong electric field enhancement that is beneficial to enhance optical nonlinearities \cite{carletti2018giant,koshelev2020subwavelength}.
For these reasons, we investigate the MS response to circularly polarized (CP) light at normal incidence when $\alpha=\alpha^*$.\\
\indent We solve the electromagnetic problem for different values of $L$, both for LCP and RCP light, in order to analyze the CD.
We use the following convention to define the CD of a certain variable $A$ \cite{chen2016giant,tang2020nano}:
\begin{equation}
    CD=\frac{A_{LCP}-A_{RCP}}{A_{LCP}+A_{RCP}}.
    \label{eq:Linear_CD}
\end{equation}
\begin{figure*}[t]
    \centering
    \includegraphics[width=1\textwidth]{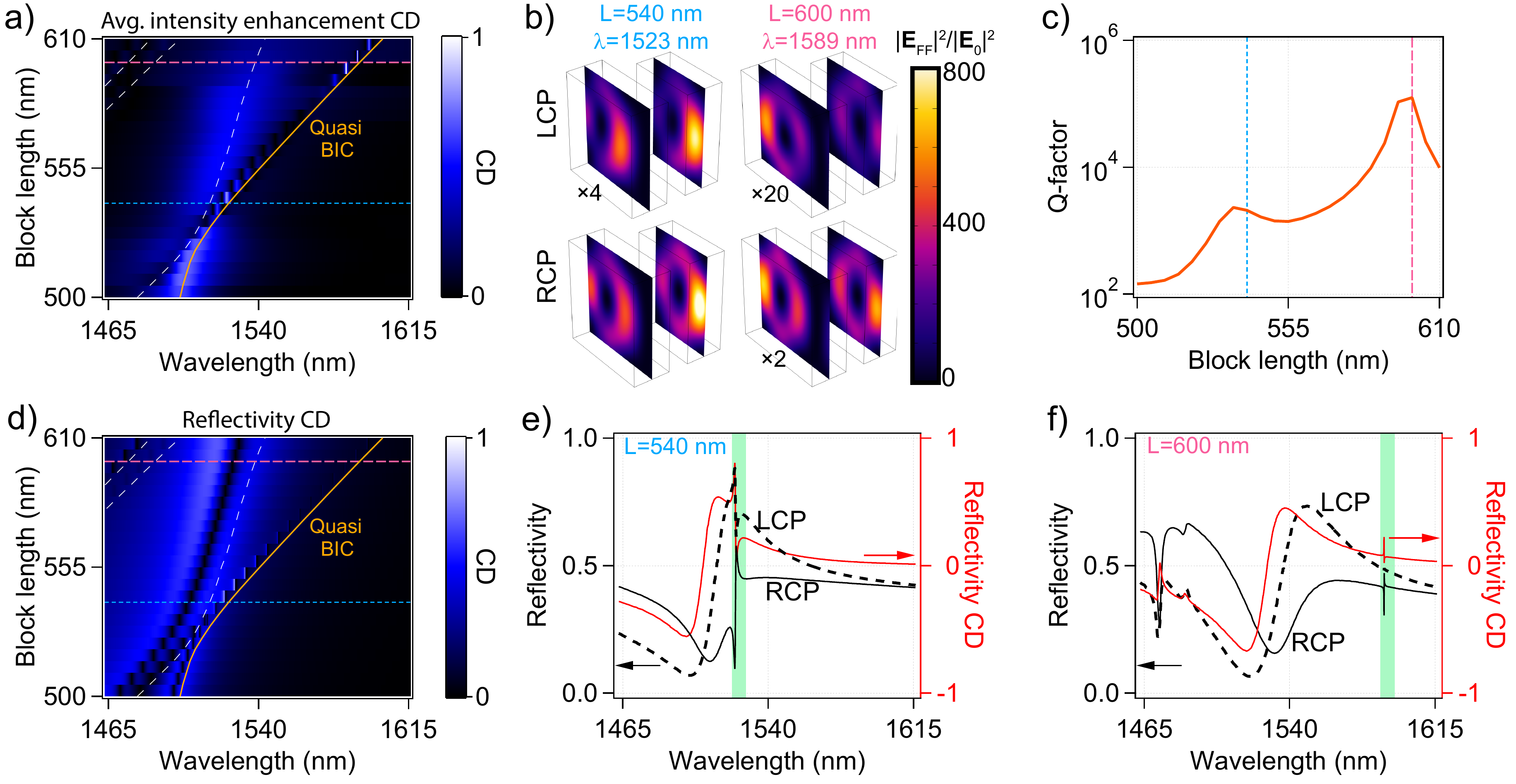}
    \caption{
    a) Average intensity enhancement CD vs block length (vertical axis) and wavelength (horizontal axis). The computed eigenmodes are reported as dashed white lines, while the quasi-BIC is highlighted by the full orange line. Horizontal cyan and red dashed lines indicate $L=540\,\text{nm}$ and $L=600\,\text{nm}$, respectively.
    b) Intensity enhancement in the two Si blocks at the quasi-BIC wavelength for LCP (top row) and RCP (bottom row) pump and for $L=540\,\text{nm}$ (left column) and $L=600\,\text{nm}$ (right column).
    The intensity enhancement of top left, top right and bottom right panels are multiplied by a factor 4, 20 and 2, respectively.
    c) $Q$-factor of the quasi-BIC vs block length.
    d) Same as a), but for the reflectivity CD.
    e) Reflectivity spectra (left axis, black) for LCP (dashed black line) and RCP (full black line) light. 
    The red line (right axis) shows the reflectivity CD. The vertical green bar highlights the wavelength of the quasi-BIC.
    The block length is 540 nm.
    f) same as e), but for $L=600$ nm.
    In all the panels, the asymmetry parameter is $0.25$.
    }
    \label{fig:CD_map}
\end{figure*}
\indent The first variable for which we calculate the CD is the average intensity enhancement, defined as:
\begin{equation}
    <I_{en}>=\frac{1}{V}\int_V \frac{I(\mathbf{r})}{I_0}dV.
    \label{Eq:av_int_enh}
\end{equation}
where $I(\mathbf{r})$ and $I_0$ are the local and the incident light intensity, and $V$ is the volume of the two Si blocks.
Fig. \ref{fig:CD_map}a shows the average intensity enhancement CD as a function of $L$ and wavelength.
A high average intensity enhancement CD can be observed when the wavelength of the incident light is tuned in proximity of a MS mode.
In particular a strong average intensity enhancement CD up to $\sim 1$ is observed close to the quasi-BIC wavelength (orange line) for $L=540\,\text{nm}$ and $L=600\,\text{nm}$ (cyan and magenta dashed lines).
The strong average intensity enhancement CD occurring at the quasi-BIC for $L=$ 540 and 600 nm can also be observed in Fig. \ref{fig:CD_map}b, where the map of $|\mathbf{E}_{FF}|^2/|\mathbf{E}_{0}|^2$ inside the Si blocks is reported ($\mathbf{E}_{0}$ and $\mathbf{E}_{FF}$ are the incident and the local electric field at the fundamental frequency, respectively).
We can observe that the quasi-BIC is selectively excited by RCP incident light and only weakly excited by LCP light. Furthermore, the high maximum intensity enhancement (up to 800) is due to the high quasi-BIC $Q$-factor that is shown as a function of the block length in Fig. \ref{fig:CD_map}c.\\
\indent Scenarios with high average intensity enhancement CD and high intensity enhancements are promising for the achievement of nonlinear CD in THG with high conversion efficiency.
Indeed, if the intensity enhancement within the Si blocks is very different upon LCP or RCP excitation, then we expect that the THG may follow the same trend.\\
\indent Fig. \ref{fig:CD_map}d shows the reflectivity CD ($R_{CD}$) spectra as a function of block length, $L$. We can observe that $R_{CD}$ around the quasi-BIC is high for $L=540$ nm, but not for $L=600$ nm.
To better appreciate this aspect, we plot the reflectivity curves (left axes, black) as a function of the wavelength (horizontal axes) for $L=540$ nm (Fig. \ref{fig:CD_map}e) and for $L=600$ nm (Fig. \ref{fig:CD_map}f).
The dashed and solid black lines correspond to LCP and RCP light illumination, respectively.
The reflectivity CD is reported as a red line (right axis, red).
As we can see, in proximity of the quasi-BIC (green vertical bar), the reflectivity CD is much higher in the case of $L=540$ nm ($R_{CD}\approx 0.8$) than for $L=600$ nm ($R_{CD}<0.25$).
The high CD obtained for the MS with $L=540$ nm is due to interference between the quasi-BIC and the other mode of the MS in the same wavelength range.
Indeed, by analysing the field distributions of these modes, we can observe that while the quasi-BIC originates from a magnetic resonance, the second resonance corresponds to an electric resonance (see Fig. \ref{fig:NL_S2} of Appendix).
This enhancement of both electric and magnetic fields provides an enhancement of the electromagnetic density of chirality \cite{solomon2018enantiospecific}.\\
\begin{figure*}[t]
    \centering
    \includegraphics[width=1\textwidth]{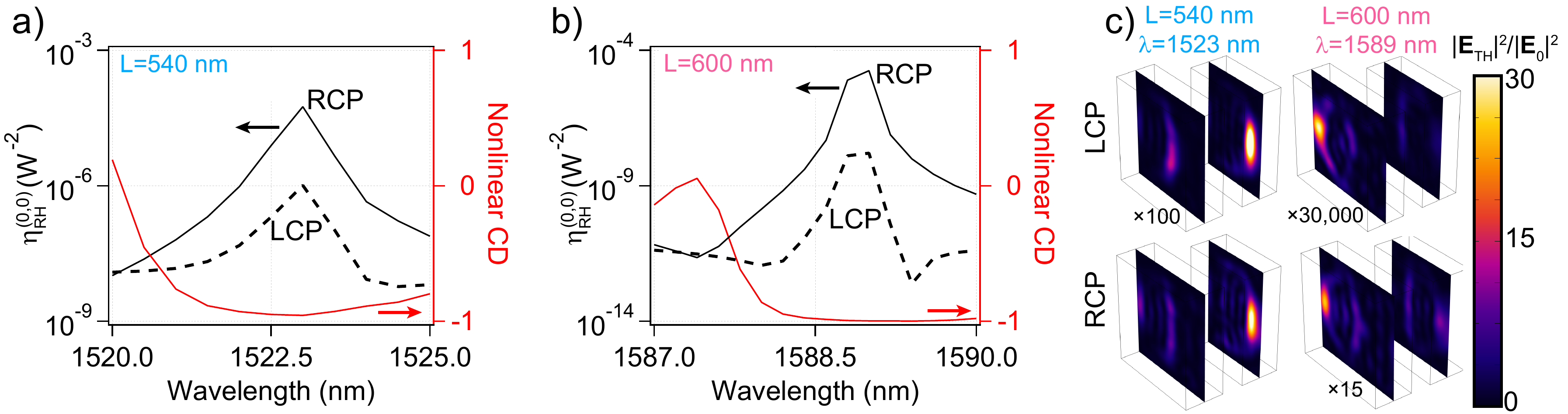}
    \caption{ a) $\eta_{RH}^{(0,0)}$ (left axis, black) vs pump wavelength for LCP (dashed black line) and RCP (full black line) pump. 
    The red line (right axis) indicates the THG CD.
    The block length is 540 nm.
    b) same as a) but with block length of $600\,\text{nm}$.
    c) $|\mathbf{E}_{TH}|^2/|\mathbf{E}_{0}|^2$ in the two Si blocks at the quasi-BIC wavelength for LCP (top row) and RCP (bottom row) pump and for $L=540\,\text{nm}$ (left column) and $L=600\,\text{nm}$ (right column).
    $|\mathbf{E}_{TH}|^2/|\mathbf{E}_{0}|^2$ has been multiplied by a factor 100 (top left), $3\times10^4$ (top right) and 15 (bottom right).}
    \label{fig:NL}
\end{figure*}
We point out that the linear CD for the transmittivity shows the same trend as $R_{CD}$.\\
\indent We analyze the THG of the MS using full-vectorial numerical simulations assuming an undepleted pump regime \cite{smirnova2016multipolar} (see Appendix for further details). 
We tune the pump wavelength in proximity of the quasi-BIC and evaluate the normalized TH conversion efficiency $\eta=P_{TH}/P_{0}^3$, where $P_{TH}$ and $P_0$ are the power at the frequency $3\omega$ emitted by the MS (in all diffraction modes and towards both air and substrate) and the incident pump power, respectively.
The conversion efficiency is maximum when  the pump  is  tuned  at the quasi-BIC wavelength regardless of the incident polarization handedness.
The $\eta$ peak value is $\sim 10^{-2}$ W$^{-2}$ (see Figs. \ref{fig:NL_S1} c and d of Appendix). This is comparable with the measured THG efficiency of hybrid multiple quantum-well and Gammadion-type plasmonic chiral nanoresonators \cite{kim2020giant}, and higher than state-of-the-art demonstration in all-dielectric structures \cite{liu2019high}.\\
\indent The wavelength
of the TH occurs in the absorption band of the silicon \cite{Green2008}. 
As a consequence, resonant modes in that range would be spectrally broad and are not expected to significantly contribute to the enhancement of the emitted TH signal.
This is also confirmed from the fact that the peak THG efficiency is obtained for a pump wavelength precisely corresponding to the quasi-BIC and that, as the pump wavelength is changed close to the quasi-BIC condition, the THG efficiency follows the quasi-BIC spectral signature.
If a mode at the TH wavelength significantly contributed to the enhancement of the THG efficiency, the peak of the latter would more likely occur at a slightly different light wavelength and result in a spectrally broader peak. This is not the case, confirming that the resonance at the pump wavelength is the leading mechanism to achieve a high THG conversion efficiency.\\
The THG emission is distributed over 9 diffraction orders since the TH wavelength is shorter than the MS period, $P$.
We focus our attention on the THG emitted in the (0,0) diffraction order in air. This light is emitted normally to the MS plane and can be collected by using low-cost optics with small numerical aperture.
By defining $P_{RH}^{(0,0)}$ and $P_{LH}^{(0,0)}$ the THG emitted powers with right-hand, RH, or left-hand, LH, polarization, we can define the polarization-dependent TH conversion efficiency through the (0,0) diffraction order as $\eta_{RH}^{(0,0)}=P_{RH}^{(0,0)}/P_0$ for RH light and $\eta_{LH}^{(0,0)}=P_{LH}^{(0,0)}/P_0$ for LH light, where $P_0$ is the incident pump power.
In Figs. \ref{fig:NL}a and \ref{fig:NL}b we report $\eta_{RH}^{(0,0)}$ (left axes, black, log scale) as a function of the incident wavelength for $L=540$ and $L=600$ nm, respectively.
In particular, the full black (dashed black) lines are obtained for a RCP (LCP) pump.\\
The strong differences between $\eta_{RH}^{(0,0)}$ obtained with RCP vs LCP pump imply the presence of an important THG CD.
The latter is plotted in Figs. \ref{fig:NL}a and \ref{fig:NL}b as red lines (right axes, red).
As we can see, in both cases the THG CD is very high in absolute value, up to 1.
Indeed, $\eta_{RH}^{(0,0)}$ may differ by more than 2 orders of magnitude if the pump is LCP or RCP.
This is due to a selective excitation of the quasi-BIC with RCP incident light with respect to LCP as observed in Fig. \ref{fig:CD_map}b.
Consequently, the TH fields in the MS are mainly focused in the region corresponding to high intensity enhancement at the quasi-BIC wavelength (see Fig. \ref{fig:CD_map}b) and increase considerably depending on the polarization handedness as it can be observed in Fig. \ref{fig:NL}c, where the map of $|\mathbf{E}_{TH}|^2/|\mathbf{E}_{0}|^2$ within the Si blocks is reported ($\mathbf{E}_{TH}$ being the electric field at frequency $3\omega$).\\
We point out that these results are obtained for RCP TH light emitted in (0,0) diffraction order towards the air.
However, identical results in terms of nonlinear CD are observed for any other diffraction order (including the ones towards the substrate) and also for LCP THG (in Figs. \ref{fig:NL_S1} a and b of Appendix we report the plot of $\eta_{LH}^{(0,0)}$ and its CD).
The same conclusions extend to the TH conversion efficiency summed over all the diffraction orders $\eta$ (see Figs. \ref{fig:NL_S1} c and d of Appendix).\\
\indent In conclusion, we have used extensive numerical simulations to characterize the linear and nonlinear chiro-optical response of a Si MS with a broken-symmetry supporting a quasi-BIC with high ($>10^5$) $Q$-factor.
Exploiting the consequent strong electric field confinement, we demonstrate a THG conversion efficiency up to $10^{-2}$ W$^{-2}$.
Owing to the chiral broken-symmetry and the mode interaction in the MS, selective excitation of the quasi-BIC is achieved for RCP incident light. This results in near-unity nonlinear CD (or THG CD) with high conversion efficiency that is comparable to state-of-the-art hybrid multi-quantum-well structures demonstrated in the infrared \cite{kim2020giant}. 
Furthermore, by tuning the length of the blocks we observe that the linear CD in reflectivity can be enhanced due to multi-mode interference quite independently from the THG CD.
Indeed, for block length $L=540$ nm both the reflectivity and THD CDs are high ($>0.8$) for incident light wavelength at quasi-BIC. On the other hand, for $L=600$ nm and at the quasi-BIC wavelength only the THG CD is high ($\sim 1$), whereas the reflectivity CD does not exceed 0.25.
\\
\indent These results show the potential of lossless all-dielectric MSs for nonlinear CD and pave the way for new applications requiring a precise control of the spin of the light such as telecommunications, quantum optics, holography, information storage, and biology.\\

\begin{acknowledgements}
This work is partially funded by the European Commission Horizon 2020 H2020-FETOPEN-2018-2020 grant agreement no. 899673 (METAFAST), by the National Research Council Joint Laboratories program, project SAC.AD002.026 (OMEN), and by the Italian Ministry of University and Research
(MIUR) through the PRIN project NOMEN (2017MP7F8F).
\end{acknowledgements}

\section*{Appendix}
\renewcommand{\thefigure}{A\arabic{figure}}

\setcounter{figure}{0}
\emph{Modelling}. The numerical models are developed in COMSOL Multiphysics. The refractive index for the Si blocks is taken from Ref. \cite{Green2008}, while the refractive index for the silica substrate is $\sim1.45$ for all the considered wavelengths \cite{malitson1965interspecimen}.\\
The computational cell is a block corresponding to the unit cell of the MS.
Periodic boundary conditions are enforced, in order to mimic the MS.
The air and substrate domains are truncated with perfectly matched layers (PML) domains to avoid spurious reflections \cite{gandolfi2018ultrafast}.\\

\emph{Eigenmode analysis}. An eigenfrequency solver is used to find the MS modes in the telecom region. The refractive index dispersion of Si is accounted by using an auxiliary-field formulation \cite{Yan2018}. The Si permittivity from Ref. \cite{Green2008} at wavelengths between $1200\,\text{nm}$ and $2000\,\text{nm}$  is modeled with a single pole Lorentz-Drude model 
\begin{equation*}
    \varepsilon(\omega)=\varepsilon_\infty-\varepsilon_\infty\frac{\omega_p^2}{\omega^2-\omega_0^2+i\omega\gamma}
\end{equation*}
with $\varepsilon_\infty=1$, $\omega_0=5.92\times10^{15}\,\text{Hz}$, $\omega_p=1.93\times10^{16}\,\text{Hz}$, and $\gamma =0\,\text{Hz}$.\\

\emph{Linear simulations}. The reflectivity and transmittivity spectra are obtained using two ports at the air-PML and substrate-PML boundaries. The excitation is set only on the former port by setting either a linearly or a circularly polarized plane wave.
A frequency domain solver is used to compute the fields at the same frequency of the incident radiation.\\

\emph{THG simulations}.
The electric fields $\mathbf{E}(\omega)$ calculated with the linear simulations are used to calculate the nonlinear polarization in the Si blocks, defined as:

$$  P_j^{NL}(3\omega)=\varepsilon_0\left\{3\chi^3_{1122}[\mathbf{E}(\omega)\cdot\mathbf{E}(\omega)]E_j(\omega)\right.+$$
$$  
    \left.+(\chi^3_{1111}-3\chi^3_{1122})E_j^3(\omega)\right\}, \ \ \ \ j=1,2,3,
    \label{Eq:polarization_3omega}$$
where $\chi^3_{1122}=2.8\times10^{-19}$ m$^2$/V$^2$ and 
$\chi^3_{1111}/\chi^3_{1122}=2.14$ \cite{moss1990band,smirnova2016multipolar}.\\
The nonlinear polarization is used as a source term for the TH calculations and
no other sources are present (i.e. there are no ports).
A frequency domain study is used to compute the fields at frequency $3\omega$. The power, $P_{TH}$, emitted by the MS at frequency $3\omega$ is:
$$ P_{TH}=\int_{\sigma}\mathbf{S}_{NL}(3\omega)\cdot\mathbf{N}d\sigma,$$
where $\mathbf{S}_{NL}(3\omega)$ is the Poynting vector at frequency $3\omega$, $\sigma$ is the external surface of the air and substrate domains (i.e. the surface in contact with the PML domains), and
$\mathbf{N}$ is the unit vector normal to the surface $\sigma$ pointing outside the computational domain (i.e. in the opposite direction with respect to the MS).\\

\begin{figure*}[t]
    \centering
    \includegraphics[width=0.68\textwidth]{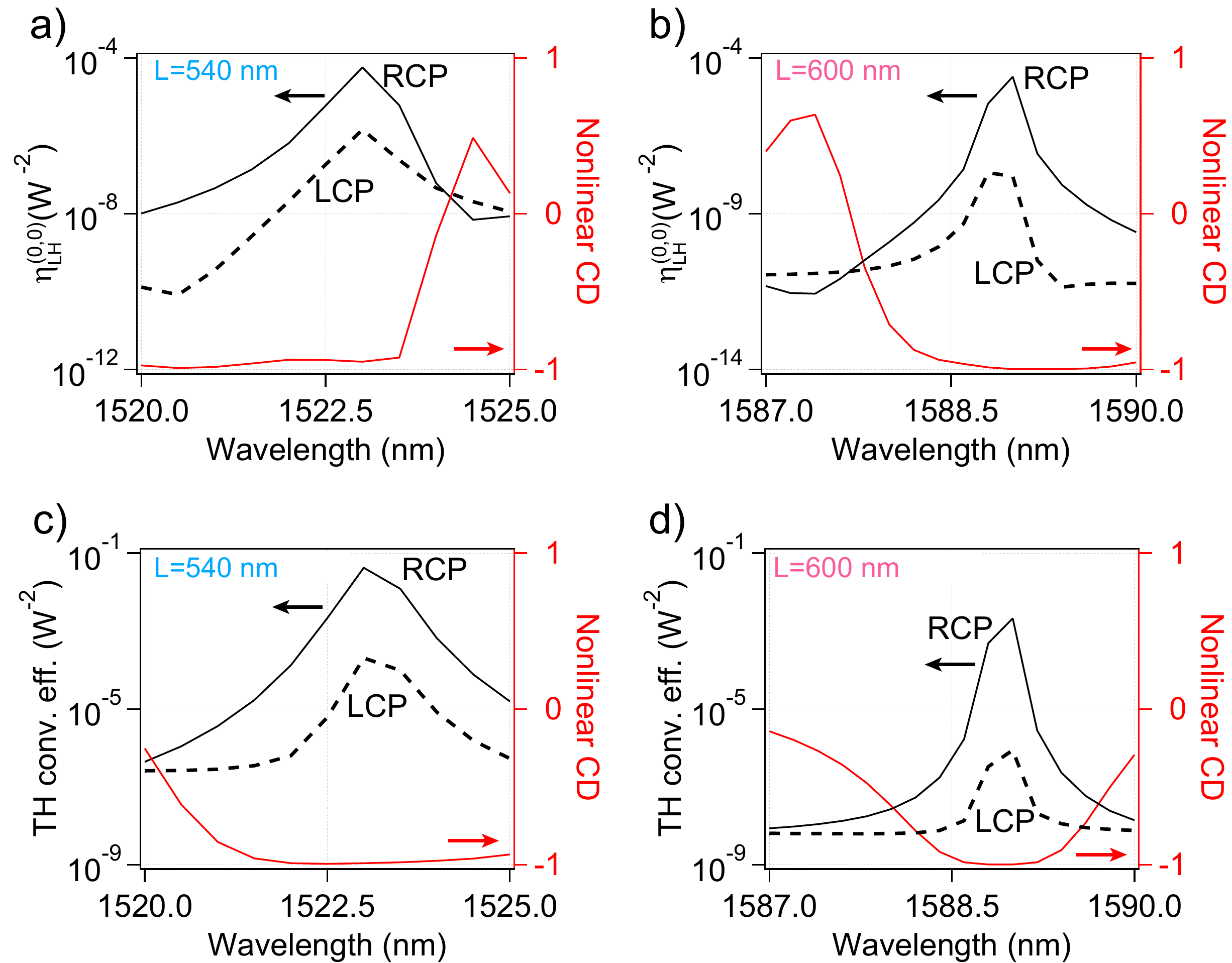}
    \caption{a) $\eta_{LH}^{(0,0)}$ and c) $\eta$ (left axis, black) vs pump wavelength for LCP (dashed line) and RCP (full line) pump. 
    The red line (right axis) indicates the THG CD.
    The block length is 540 nm.
    b) and d) same as a) and c) respectively, but with block length of $600\,\text{nm}$.}
    \label{fig:NL_S1}
\end{figure*}

\emph{CD for LH light emitted through the diffraction order (0,0) towards the air.}
In Fig. \ref{fig:NL_S1} a) and b) we report $\eta_{LH}^{(0,0)}$ and its CD, as functions of the wavelength, in analogy to Fig. \ref{fig:NL}.\\

\emph{Normalized total THG conversion efficiency}. In Fig. \ref{fig:NL_S1} c) and d) we report the total TH conversion efficiency $\eta=P_{TH}/P_{0}^3$ vs pump wavelength, where $P_{TH}$ is the power at the frequency $3\omega$ emitted by the MS in all diffraction orders, towards both air and substrate and with both LCP and RCP polarizations.
The full black (dashed black) line is obtained for RCP (LCP) pump.
$P_0$ is the incident pump power. The corresponding nonlinear CD is reported as red lines (right axes, red).\\

\emph{$Q$-factor reduction due to fabrication defects}.
The results reported in the manuscript are obtained for an ideal structure, with structures with vertical and smooth side-walls and infinite, exactly identical unit cells.
A fabricated device, provided that the patterned area is large enough to accommodate some tens of unit cells, will differ from this ideal structure mainly in both side-wall quality (verticality and roughness) and lattice uniformity. These fabrication imperfections and defects will decrease the photons lifetime in the mode.
To account for these phenomena, we can define a photon lifetime $\tau_{def}$ associated with fabrication features.
Hence, the total photons lifetime, $\tau_{tot}$, may be expressed as
$$\frac{1}{\tau_{tot}} =\frac{1}{\tau}+\frac{1}{\tau_{def}}.$$
where $\tau$ is the photon lifetime of the ideal structure.
Thus, the lifetime $\tau_{def}$ strongly depends on the specific fabrication procedure used for the realization of the metasurface.\\
The $Q$-factor is equal to $Q_{tot}=\omega_0 \tau_{tot}$ with $\omega_0$ the angular frequency of the quasi-BIC.
The photons lifetime due to fabrication related features decreases the $Q$-factor according to 
$$\frac{1}{Q_{tot}} =\frac{1}{Q}+\frac{1}{Q_{def}}.$$
Comparing the measured $Q$-factor with simulations results, an estimation of $\tau_{def}$ lifetime can be obtained.
\\

\begin{figure*}[t]
    \centering
    \includegraphics[width=0.4\textwidth]{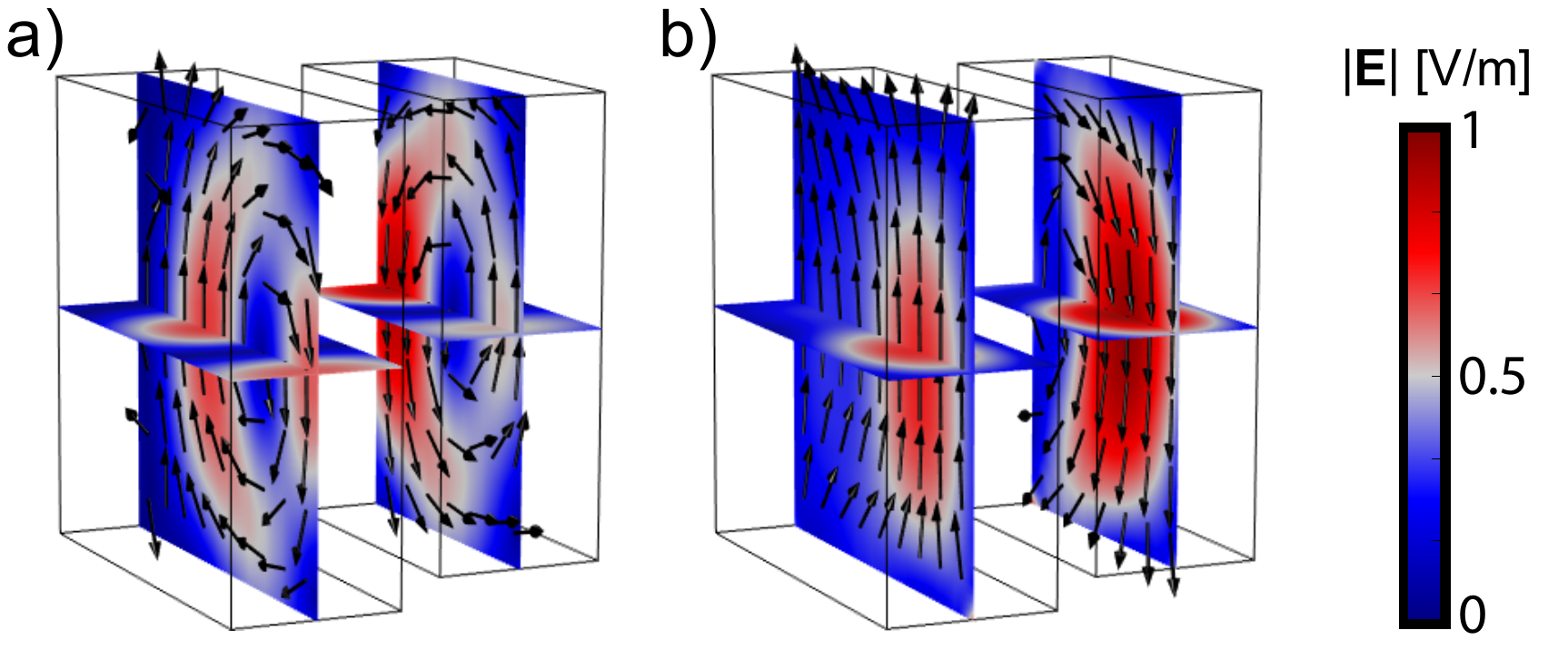}
    \caption{Normalized electric field profiles for the quasi-BIC (a) and closer mode (b) for a structure with $L=540$ nm and $\alpha=0.25$. The electric field vector is shown with black arrows.}
    \label{fig:NL_S2}
\end{figure*}

\emph{Electric field profiles of the quasi-BIC and the closer mode for the structure with $L=540$ nm}.
In Fig. \ref{fig:NL_S2}a we report the electric field profile of the quasi-BIC mode of the structure with $L=540$ nm and $\alpha=0.25$, occurring at wavelength 1525 nm.
The mode profile shows a magnetic resonance.\\
In Fig. \ref{fig:NL_S2}b we display the electric field profile of the other mode, at wavelength 1515 nm, falling closer to the quasi-BIC.
This second mode corresponds to an electric resonance.\\

Marco Gandolfi, Andrea Tognazzi, Davide Rocco, Costantino De Angelis, and Luca Carletti
Phys. Rev. A 104, 023524.\\
Copyright 2021 by the American Physical Society.
\bibliography{Gandolfi_references}

\end{document}